\documentclass[journal=jacsat,manuscript=article]{achemso}
\usepackage[version=3]{mhchem} 

\usepackage{xcolor}
\usepackage{hyperref}
\usepackage{physics}
\usepackage{soul}

\author{Mohit Gupta}
\affiliation[UMN]{School of Physics and Astronomy, University of Minnesota, Minneapolis, Minnesota 55455, USA}

\author{Vipin Khade}
\affiliation[UMN]{School of Physics and Astronomy, University of Minnesota, Minneapolis, Minnesota 55455, USA}

\author{Colin Riggert}
\affiliation[UMN]{School of Physics and Astronomy, University of Minnesota, Minneapolis, Minnesota 55455, USA}

\author{Lior Shani}
\affiliation[UMN]{School of Physics and Astronomy, University of Minnesota, Minneapolis, Minnesota 55455, USA}

\author{Gavin Menning}
\affiliation[UMN]{School of Physics and Astronomy, University of Minnesota, Minneapolis, Minnesota 55455, USA}

\author{Pim J. H. Lueb}
\affiliation[TUE]{Department of Applied Physics, Eindhoven University of Technology, Eindhoven 5600 MB, The Netherlands}

\author{Jason Jung}
\affiliation[TUE]{Department of Applied Physics, Eindhoven University of Technology, Eindhoven 5600 MB, The Netherlands}

\author{R\'egis M\'elin}
\affiliation[ING]{Universit\'e Grenoble-Alpes, CNRS, Grenoble INP, Institut NEEL, Grenoble 38042, France}

\author{Erik P. A. M. Bakkers}
\affiliation[TUE]{Department of Applied Physics, Eindhoven University of Technology, Eindhoven 5600 MB, The Netherlands}

\author{Vlad S. Pribiag}
\affiliation[UMN]{School of Physics and Astronomy, University of Minnesota, Minneapolis, Minnesota 55455, USA}
\email{vpribiag@umn.edu}

\title[An \textsf{achemso} demo]{Evidence for \texorpdfstring{$\pi$}--shifted Cooper quartets and few-mode transport in PbTe nanowire three-terminal Josephson junctions}

\abbreviations{IR,NMR,UV}

\begin{document}

\begin{abstract}
Josephson junctions are typically characterized by a single phase difference across two superconductors. This conventional two-terminal Josephson junction can be generalized to a multi-terminal device where the Josephson energy contains terms with contributions from multiple independent phase variables. Such multi-terminal Josephson junctions (MTJJs) are being considered as platforms for engineering effective Hamiltonians with non-trivial topologies, such as Weyl crossings and higher-order Chern numbers. These prospects rely on the ability to create MTJJs with non-classical multi-terminal couplings in which only a few quantum modes are populated. Here, we demonstrate these requirements in a three-terminal Josephson junction fabricated on selective-area-grown (SAG) PbTe nanowires. We observe signatures of a $\pi$-shifted Josephson effect, consistent with inter-terminal couplings mediated by four-particle quantum states called Cooper quartets. We further observe supercurrent co-existent with a non-monotonic evolution of the conductance with gate voltage, indicating transport mediated by a few quantum modes in both two- and three-terminal devices.
\end{abstract}

\maketitle
Hybrid superconductor-semiconductor Josephson devices offer a promising path to engineering tunable quantum matter because they combine intrinsic quantum correlations and coherence (stemming from the superconductor) with precise local electrostatic control of the Josephson coupling, number of quantum modes and scattering properties (owing to their simultaneous semiconducting character). Superconductivity across the semiconductor is mediated by Andreev bound states (ABS)~\cite{Meissner1960, Josephson1962, Anderson1963}, which result from the coherent Andreev reflections at the superconductor-semiconductor interfaces. In general, the Josephson energy of a junction with $N$-terminals contains $N-1$ independent phase differences. The ABS of such multi-terminal Josephson junctions (MTJJs) can theoretically provide a means to realize and tune Bloch-like Hamiltonians with non-trivial topology, including Weyl crossings and higher-order Chern numbers~\cite{Riwar2016, Meyer2017,Xie2017,Xie2018,Xie2022}. However, the realization of these effective Hamiltonians requires multiple superconducting electrodes to be non-classically coupled across a single scattering region and the number of conductance modes connecting the superconducting terminals must be close to unity~\cite{Riwar2016,Meyer2017,Eriksson2017}, which has to date proven challenging.

Recent attempts to realize gate-tunable MTJJs have focused on two-dimensional electron systems, such as graphene or III-V semiconductor quantum wells (e.g. InAs)~\cite{Draelos2018,Pankratova2020,Gino2020,arnault2021multiterminal,graziano2022,Arnault2022,Gupta2023}. In these devices, the multi-terminal couplings are negligible compared to the two-terminal couplings and the transport properties can thus be explained by a classical network of two-terminal Josephson junctions. The absence of non-classical multi-terminal couplings in these studies is likely due to the extended planar geometries imposed by the two-dimensional materials and the resulting absence of a compact central scattering region. Due to the absence of non-classical inter-terminal couplings, such device geometries are unlikely to realize multi-terminal ABS and topological Hamiltonians. To address this challenge, here we investigate MTJJs based on semiconductor nanowires, which cross in a single central region. The semiconductor platform consists of selective-area-grown (SAG) PbTe nanowires. The nanowires can be grown along multiple crystallographic directions~\cite{Fabrizio_Bakkers2022} and allow for the nanofabrication of MTJJs where the natural quantum confinement resulting from the wire cross-section at the point of intersection facilitates coherent coupling of multiple superconducting electrodes in a central scattering region. 

It was recently theorized that harmonic contributions with $\pi$-shifted phase should be present in the current-phase relation of MTJJs with non-classical inter-terminal couplings. This $\pi$-shift results from phase-coherent four-particle processes involving all three terminals, called Cooper quartets ~\cite{Freyn2011,Melin2023,Melin2023_3tjj,Jonckheere2023,melin2024magnetointerferometry}. We observe signatures of this $\pi$-shifted harmonic in the diffraction pattern, in the form of an enhanced critical current for small magnetic fields~\cite{bulaevskii1978,weides2006,frolov2006,Kang2022}, along with characteristic differential resistance maps in the space of bias currents. Our results open up an efficient and promising new path to fabricate compact multi-terminal Josephson devices with an arbitrary number of terminals and tunable couplings. These devices may have applications ranging from constructing topologically nontrivial Hamiltonians in $N$-dimensional phase space~\cite{Riwar2016,Meyer2017,Xie2017,Xie2018,Xie2022} to multi-signal intermodulation which can be used as a building block for neuromorphic computing~\cite{Goteti2021,Gupta2023}.

Semiconductor nanowires coupled with superconductors are also a leading candidate system for studying Majorana zero modes (MZMs)~\cite{Lutchyn2010,Oreg2010, Lutchyn2018}, which are predicted non-Abelian states~\cite{Kitaev2003,Nayak2008}. One of the leading challenges toward the realization of MZMs in nanowire devices to date has been disorder~\cite{Liu2012,Mourik2012,Pan2021,Pan2021_2,Ahn2021,jiang2023zerobias,frolov2023smoking}. Recently, PbTe semiconducting nanowires have gained attention as an alternative material platform to realize MZMs, with the expectation that the large dielectric constant of PbTe ($\sim1000$ in bulk) may help screen disorder~\cite{Huang2021,Cao2022}. PbTe nanowires have the other necessary ingredients for realizing MZMs, such as large g-factors and spin-orbit coupling~\cite{Fabrizio_Bakkers_gfactor_2022,Fabrizio_Bakkers2022,Gomanko2022}. Recently, ballistic one-dimensional transport ~\cite{Song2023,Wang2023,song2024reducing} and induced superconductivity~\cite{HaoZhang2023_SC,li2023selectiveareagrown} have been achieved in PbTe SAG nanowires. However, demonstration of few mode transport coexistent with induced superconductivity is lacking. In our work, we observe plateau-like conductance features co-existent with supercurrent in both two- and three-terminal devices, indicating superconductivity in the regime of a few quantum modes. 

\begin{figure}[!h]
    \includegraphics[width=1\textwidth]{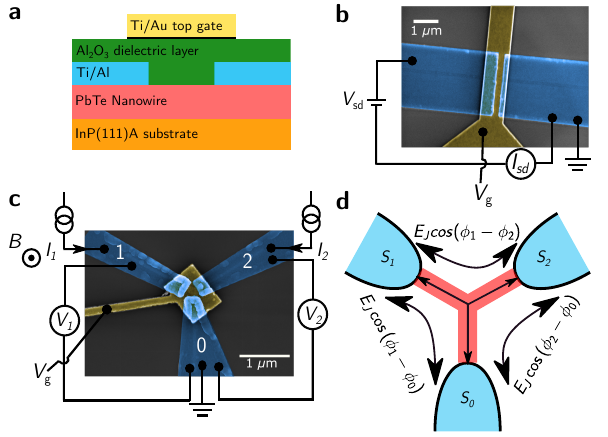}
    \caption{\textbf{Device schematics.} \textbf{a,} Cross-sectional schematic of the device material stack (not to scale).  \textbf{b,} False-color scanning electron microscope (SEM) image of a two-terminal Josephson junction.  \textbf{c,} SEM image of a three-terminal MTJJ (Device 4) along with measurement schematics used for the measurements on Device 3. The blue region corresponds to the superconducting electrodes and gold with the electrostatic gate. PbTe wires are visible underneath the Al electrodes. \textbf{d,} Schematic of transport in a three-terminal MTJJ, showing Josephson couplings between pairs of contacts. Terms corresponding to multi-terminal couplings are not displayed.}
    \label{sem_dev}
\end{figure}

The PbTe nanowire synthesis is discussed in Supporting Information Section I. Device contacts are deposited by evaporating either Ti/Al (superconducting contacts) or Ti/Au (normal contacts). Electrostatic control is enabled by metallic top gates with AlO$_{\rm{x}}$ gate dielectric, resulting in the stack structure shown in Figure~\ref{sem_dev}\textbf{a}. Scanning electron microscope (SEM) images of two- and three-terminal Josephson junctions (3TJJs), along with measurement schematics, are shown in Figure~\ref{sem_dev}\textbf{b} and \textbf{c} respectively. Two-terminal devices are measured in the voltage-bias configuration and three-terminal devices are measured in the current-bias configuration unless otherwise specified. The spacing between superconducting electrodes is measured to be $\sim 150$ nm for two-terminal devices and $\sim 250$ nm for three-terminal devices. We present four devices in this work, labeled Device $1-4$. 

\begin{figure}[!h]
    \includegraphics[width=\textwidth]{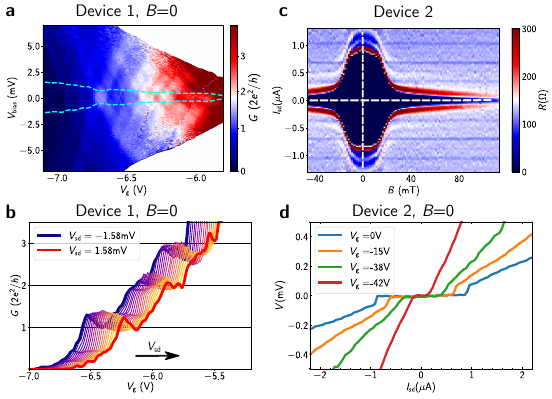}
    \caption{\textbf{Ballistic transport characteristics in two-terminal devices.} \textbf{a,} Conductance as a function of $V_{\rm{bias}}$ and $V_{\rm{g}}$ for Device 1 at $B=0$. \textbf{b,} Conductance as a function of $V_{\rm{g}}$ for different $V_{\rm{sd}}$ for Device 1 at $B=0$. The curves correspond to $V_{\rm{sd}}$ values between $-1.58$ mV and $1.58$ mV (shown by dashed cyan lines in panel \textbf{a}) in increments of $0.150$ mV. The curves are each offset along the $V_{\rm{g}}$ axis for clarity with an arrow indicating the direction of increasing $V_{\rm{sd}}$. \textbf{c,} $R$ as a function of $I_{\rm{sd}}$ and $B$ for Device 2, dashed white lines show zero of $B$ and $I_{\rm{sd}}$. \textbf{d,} $I-V$ characteristics for different values of $V_{\rm{g}}$ for Device 2.}
    \label{Iv_char}
\end{figure}

Device 1 uses Ti/Au contacts (see Supporting Figure S2\textbf{a} for the device image). The device is measured in a voltage bias configuration of the type shown in Figure~\ref{sem_dev}\textbf{b}, in a dilution refrigerator, at the base temperature of $\sim 8$ mK. Using standard lock-in techniques we characterize the differential conductance, G, of this device as a function of gate voltage $V_{\rm{g}}$ and $V_{\rm{bias}}$, where $V_{\rm{bias}}=V_{\rm{sd}}-I_{\rm{sd}}*R_{\rm{series}}$, with $V_{\rm{sd}}$ being the source-drain voltage bias, $I_{\rm{sd}}$ the measured source-drain current, and $R_{\rm{series}}$ includes the resistance of the low-pass filters, the transimpedance amplifier and the contact resistance. We observe conductance steps for the first few modes, at zero magnetic field, in this device with a 200 nm gate-defined channel. The diamonds of quasi-uniform conductance in the $V_{\rm{g}}$ vs. $V_{\rm{bias}}$ map are visible in Figure \ref{Iv_char}\textbf{a}. Linecuts at fixed $V_{\rm{sd}}$, shown in Figure \ref{Iv_char}\textbf{b}, confirm these to correspond to the 1, 2, and 3 $G_0$  quantized conductance plateaus associated with the first three spin-degenerate subbands (with $G_0=2e^2/h$, where $e$ is the elementary charge and $h$ the Planck constant). This zero-field conductance quantization is consistent with recent results published on PbTe SAG wires grown on CdTe \cite{Song2023,Wang2023,song2024reducing}. Zero-field quantized conductance, as seen in this device, is an important milestone towards realizing MZMs in nanowires, as the transition to topological superconductivity requires few-mode ballistic transport and fields of only a few hundreds of mT \cite{Mourik2012,Lutchyn2010,Oreg2010}. Additionally, zero-field quantized conductance is qualitatively suggestive of low levels of device disorder\cite{vanWeperen2013}, the importance of which has been stressed in recent theoretical studies of MZMs in realistic device architectures, which suggest that disorder is the largest impediment in the clear observation of MZMs in nanowires \cite{Pan2021,Pan2021_2,Lutchyn2010}. 

We have also performed conductance measurements at high magnetic fields and elevated temperatures, which further support the claim of conductance quantization in Device 1 (See Supporting Information section II and Supporting Figure S2). Several checkerboard-like resonances are superimposed on the conductance data (Figure \ref{Iv_char}\textbf{a} and Supporting Figure S2). These are consistent with an accidental gate-tunable quantum dot present in these nanowires~\cite{Gomanko2022}.

Having discussed the normal state transport, we now characterize the superconducting properties of a two-terminal device with Ti/Al contacts (Device 2). We first set $V_{\rm{g}}$ to a value of $0$ V. Upon subtraction of $R_{\rm{series}}$, the device differential resistance, $R=dV_{\rm{sd}}/dI_{\rm{sd}}-R_{\rm{series}}$, is obtained as a function of $I_{\rm{sd}}$ and the applied out-of-plane magnetic field, $B$ (Figure~\ref{Iv_char}\textbf{c}). The observed diffraction pattern has only one central lobe, consistent with one-dimensional supercurrent flow in this 100nm diameter wire. We also observe Fiske resonances~\cite{coon1965josephson} above the zero resistance region in the central lobe, indicating a well-defined Josephson cavity. A similar diffraction pattern is observed as a function of the in-plane magnetic field (Supporting Figure S3\textbf{a}), further confirming the one-dimensional nature of supercurrent flow. We have also measured Josephson junctions with different wire diameters, down to a minimum of $60$ nm (see Supporting Figure S3\textbf{b, c}), showing good reproducibility. Tuning $V_{\rm{g}}$ to negative values 
reduces the switching current ($I_s$) and increases the normal state resistance, showing induced gate-tunable superconductivity (Figure~\ref{Iv_char}\textbf{d}). Somewhat large electric fields (corresponding to $V_{\rm{g}}\sim -40V$) are required to observe an appreciable change in the critical current of this device. This could be due to unintentional higher levels of Pb in some of the nanowires from growth, or other unknown wire non-idealities.

\begin{figure}[!h]
    \includegraphics[width=\textwidth]{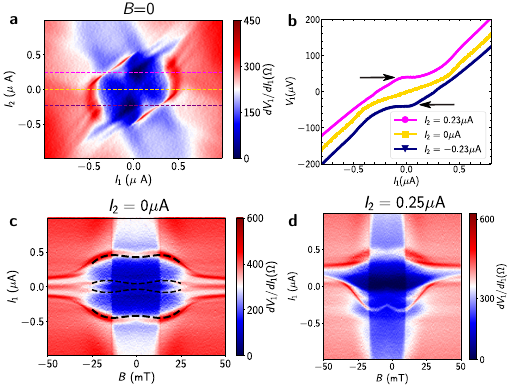}
    \caption{\textbf{Three-terminal transport} \textbf{a,} Differential resistance map as a function of bias currents $I_1$ and $I_2$ for Device 3 at $B=0$. \textbf{b,} $I_1-V_1$ characteristics for different values of $I_2$ shown by dashed lines in \textbf{a}. The curves are offset vertically for clarity by $40\mu$V, $0\mu$V and $-40\mu$V for $I_2=0.23\mu$A, $I_2=0\mu$A and $I_2=-0.23\mu$A, respectively. Arrows indicate the region of vanishing differential resistance. \textbf{c, d,} Differential resistance as a function of $B$ and bias current $I_1$ at $I_2=0$ and $I_2=0.25\mu$A, respectively. The dashed black lines in \textbf{c} show the fit to the $\pi$-shifted quartet model.}
    \label{3tjj_map}
\end{figure}

We now turn our attention to the three-terminal Josephson junctions. To map out their phase diagrams, two bias currents, $I_1$ and $I_2$, are varied, and two voltages, $V_1$ and $V_2$, are measured (see Figure~\ref{sem_dev}\textbf{c} for measurement schematic), resulting in differential resistance maps, as shown in Figure~\ref{3tjj_map}\textbf{a} for Device 3. We observe three superconducting arms in the differential resistance maps along with a central superconducting feature. The superconducting arms are roughly along $I_2\sim\frac{3}{2} I_1$, $I_2\sim-\frac{5}{2}I_1$, and $I_2\sim-\frac{1}{2}I_1$, where $V_1-V_2=0$, $V_1=0$, and $V_2=0$, respectively (see Supporting Figure S4\textbf{a} for the full dataset). In the central superconducting region, at $B=0$, superconductivity appears to be split into two distinct pockets at a finite value of $I_2$ separated by regions of finite resistance. Importantly, when both currents are set to zero, we observe a finite differential resistance as shown by the non-zero slope of $I_1-V_1$ curve at $I_2=0$ in Figure~\ref{3tjj_map}\textbf{b}. At finite $I_2$, we observe regions of constant $V_1$ (indicated by arrows in Figure~\ref{3tjj_map}\textbf{b}), yielding zero differential resistance. $V_1$ values are corrected for the offset due to the instruments and the series resistance due to the contacts. Such features in the central superconducting region have not been observed in previous MTJJ experiments. In the presence of a $0-\pi$ Josephson current-phase relation (C$\varphi$R), the critical current can be vanishing at $B=0$~\cite{bulaevskii1978,weides2006,frolov2006,Kang2022}. The observation of finite differential resistance near zero current biases with vanishing differential resistance at finite $I_2$-bias suggests the presence of $\pi$-shifted harmonic in our device, stemming from multi-terminal coherent processes~\cite{Freyn2011,Melin2023,Melin2023_3tjj,Jonckheere2023,melin2024magnetointerferometry}. 

To further clarify this point, we performed magnetic field measurements at fixed values of $I_2$, sweeping $I_1$ and varying the out-of-plane magnetic field, $B$. For $I_2=0$, the resulting interference patterns on these devices show a non-convex pattern, i.e. critical current increases with magnetic field for small $B$, as can be seen in Figure~\ref{3tjj_map}\textbf{c} in the transition from the superconducting state to the normal state. At $I_1\sim 0$, a feature with finite differential resistance is present that converges toward $I_1=0$ near $B=0$. Such non-convex diffraction patterns in two-terminal devices have been interpreted as a signature of the $0-\pi$ C$\varphi$R~\cite{bulaevskii1978,weides2006,frolov2006,Kang2022}, further suggesting the existence of $\pi$-shifted harmonic in our device. 

To understand the transport properties of the differential resistance map and $0-\pi$ C$\varphi$R, we model our device's Josephson energy including terms corresponding to both the usual two-terminal process, with energy $E_J$, and a three-terminal process, with energy $E'_J$, as shown in Equation~\ref{cpr_mtjj}. In the presence of only the two-terminal process, the device can be modeled as a semi-classical resistively and capacitively shunted junction (RCSJ) network model, as discussed in Refs.~\cite{Gino2020,graziano2022}. The superconducting arms in the phase diagram appear when a single Josephson junction present between any pair of terminals is superconducting. The slopes of these superconducting arms in the $I_1$-$I_2$ plane can be understood by the resistor network model presented in Refs.~\cite{Gino2020,graziano2022,Gupta2023}.
\begin{eqnarray} 
\begin{aligned}
    E=-E_J\cos{(\phi_1-\phi_0)}-E_J\cos{(\phi_2-\phi_0)}-E_J\cos{(\phi_1-\phi_2)}\\
    - E'_J\cos{(\phi_1+\phi_0-2\phi_2)}-          E'_J\cos{(\phi_2+\phi_0-2\phi_1)}-E'_J\cos{(\phi_1+\phi_2-2\phi_0)}
    \label{cpr_mtjj}
    \end{aligned}
\end{eqnarray}

However, the central superconducting feature and the non-convex interference pattern cannot be captured by this RCSJ network model. To fit the observed interference pattern, we derive an expression for the critical current, $I_c$ as a function of the magnetic field by starting from the expression for the total Josephson energy of the device, Equation~\ref{cpr_mtjj}. The $\pi$-shifted C$\varphi$R sets $E'_J<0$~\cite{Melin2023}. Gauge-invariance and the applied magnetic field allow us to write, $\phi_1-\phi_0=\frac{2\pi B A}{\phi_e}=\Phi$, where $B$ is the applied magnetic field, $A$ is the junction area and $\phi_e$ is the flux quantum. In our model, we have assumed that all three leads have identical couplings to the central scattering region. We arrive at the following expression for the critical current as a function of magnetic field (See Supporting Information section V for the full model derivation) for small magnetic fields and small $E'_J$:
\begin{equation}
    I_c(\Phi)=\frac{4e}{\hbar}\lvert E_J\cos(\Phi/2)+E'_J\cos(3\Phi/2)\rvert
    \label{cpr_mtjj_2}
\end{equation}

We fit Equation~\ref{cpr_mtjj_2} to the non-convex feature present at $I_1\sim 0.5\mu$A in our experimental data and obtain the curve shown in Figure~\ref{3tjj_map}\textbf{c} by a dashed black line, with $E_J=0.26\frac{\hbar}{2e}\mu$A and $E'_J=-0.05\frac{\hbar}{2e}\mu$A and $A=0.029\mu\mathrm{m^2}$. The good agreement between our experiment and theory serves as compelling evidence for the presence of quartet transport due to correlated Cooper pairs across the three superconducting terminals. The finite differential resistance feature that converges toward $I_1=0$ near $B=0$ can also be fitted with the same equation. This is fitted with $E_J=0.04\frac{\hbar}{2e}\mu$A and $E'_J=-0.02\frac{\hbar}{2e}\mu$A and $A=0.035\mu\mathrm{m^2}$. The fit is shown by thin dashed black lines converging towards $I_1=0$ in Figure~\ref{3tjj_map}\textbf{c}. The values of $A$ differ from the lithographically defined dimensions, which can be attributed to flux focusing effects~\cite{Pribiag2015,Paajaste2015,HARADA2002229}. The exact nature of the doubling of the quartet resonance is not clear at this point, and our basic theoretical model does not account for this. However, both resonances are well fitted with the derived quartet expression, suggesting a substantial three-terminal contribution, with energy in the range of 20$\%$ to 50$\%$ that of the two-terminal contribution.

Next, we study the quantum interference patterns for $I_2 \neq 0$. For positive (negative) values of $I_2$, the non-convex diffraction pattern appears only for the negative (positive) value of $I_1$ and remains flat for positive (negative) values of $I_1$ (Figure~\ref{3tjj_map}\textbf{d} and Supporting Figure S4\textbf{b}). This shows that the current bias via terminal 2 can be used as another knob for tuning the phase differences, in addition to the applied field, resulting in a superconducting diode effect. Our current model does not consider these processes, and hence, we have not fitted any theoretical curve to these data. In addition, the center of the interference pattern is shifted along the $I_1$ axis. This shift is simply a consequence of current conservation.

\begin{figure}[!h]
    \includegraphics[width=\textwidth]{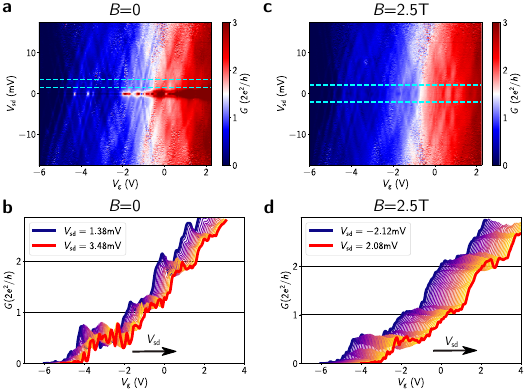}
    \caption{\textbf{Superconducting transport in the few-modes regime.} \textbf{a, c,} Conductance as a function of source-drain bias, $V_{\rm{sd}}$, and gate voltage, $V_{\rm{g}}$, for Device 4 at $B=0$T and $B=2.5$T, respectively. \textbf{b, d,} Conductance as a function of gate voltage for different $V_{\rm{sd}}$ at $B=0$T and $B=2.5$T, respectively. The curves correspond to $V_{\rm{sd}}$ values of $1.38$ mV to $3.48$ mV (panel \textbf{b}) and $-2.12$ mV to $2.08$ mV (panel \textbf{d}), shown by the dashed cyan lines in \textbf{a} and \textbf{c}. The curves are each offset along the $V_{\rm{g}}$ axis for clarity. Arrows indicate the direction of increasing $V_{\rm{sd}}$.
}
    \label{SC_QPC}
\end{figure}

The superconducting devices discussed so far required relatively large negative gate voltages to modulate conductance and could not be safely pinched off. However, from the same material growth batch, there are devices on which complete pinch-off of the channel is observed (both two- and three-terminal devices). We demonstrate few-mode transport co-existent with superconductivity on a three-terminal device (Device 4). Figure~\ref{SC_QPC}\textbf{a} shows the differential conductance, G, map for a pair of terminals as a function of $V_{\rm{sd}}$ and $V_{\rm{g}}$ for Device 4. Here, $G=1/R$ and $R=dV_{\rm{sd}}/dI_{\rm{sd}}-R_{\rm{series}}$. When the device is superconducting, the voltage drop across the device vanishes, resulting in $V_{\rm{sd}}=I_{\rm{sd}}\cross R_{\rm{series}}$. Critical current contours are thus observed as areas of high conductance or zero resistance for small bias. Plateau-like conductance features are seen at higher bias values as a function of $V_{\rm{g}}$ (Figure~\ref{SC_QPC}\textbf{b}), with superimposed conductance oscillations. The oscillations could be due to Fabry-P\'{e}rot interference or an accidental quantum dot present in the wire~\cite{Calado2015,graziano2022,shani2024diffusive}. The plateau features are not quantized in multiples of $G_0$. This is likely due to the high source-drain bias. At finite bias, the value of the conductance steps is determined by the number of quasi-1D subbands falling within the bias window set by $V_{\rm{sd}}$~\cite{Kouwenhoven1989,Patel1991,Gallagher2014,Lee2019}. It is not possible to perform zero bias conductance quantization measurements on a Josephson device as the device is superconducting at $V_{\rm{sd}}=0$. Similar results were reproduced on a two-terminal Josephson device (Supporting Figure S5). Taken together, the presence of this non-monotonic, plateau-like conductance behavior at above-gap bias values, along with supercurrent at lower bias, indicates few-mode operation of these superconducting nanowire devices. This is one of the necessary conditions for the observation of Weyl nodes in the ABS of MTJJs and the detection of MZMs in two-terminal devices. 

To further characterize the ballistic few-mode transport in Device 4, we apply a 2.5 T out-of-plane magnetic field to drive the device from the superconducting to normal state and again perform DC measurements as a function of $V_{\rm{g}}$ and $V_{\rm{sd}}$. The resulting data, plotted in Figure ~\ref{SC_QPC}\textbf{c}, shows diamond-shaped regions of approximately uniform conductance. Linecuts of this data, shown in Figure \ref{SC_QPC}\textbf{d}, show better-defined conductance plateau features with reduced conductance oscillations, attributed to the suppression of coherent backscattering due to the Aharanov-Bohm phase contribution. The plateau features become aligned with multiples of 0.5 $G_0$ at low bias, consistent with Zeeman-split subbands. This evolution to half-quantization under applied field further strengthens the interpretation of few-mode transport at zero field.

We note that out of the three pairs of terminals on this device, two pairs showed few-mode transport, while the third pair showed similar behavior to Device 2, where large electric fields were needed to pinch off the device. We also observed checkerboard conductance features superimposed on the conductance map, similar to those seen for Device 1. Such features, attributed to accidental quantum dots, are observed in all devices that show depletion of the scattering region for reasonable values of gate voltage ($V_{\rm{g}}\sim6V$). We resolve these quantum dot features in Device 4 by exploring the low conductance region (Supporting Information section VI and Supporting Figure S6). We find results consistent with previous characterizations of quantum dots on PbTe, such as near-zero charging energy and large anisotropic g-factors~\cite{Gomanko2022,Fabrizio_Bakkers2022}.

To our knowledge, there has been no prior experimental detection of $\pi$-shifted quartets. Other resonant features in the differential resistance maps of MTJJs, obtained in the non-equilibrium regime~\cite{Cohen2018,Huang2022}, can also point to quartet processes, however in such cases, the experimental features can also be due to correlated phase dynamics in a classical circuit in the RCSJ picture~\cite{graziano2022,Arnault2022}, and thus may not necessarily be due to quartets.

As further confirmation of our analysis, we consider alternative explanations that could give rise to the observed non-convex patterns, i.e. field enhanced critical currents. These include quasiparticle cooling effects~\cite{Murani2020}, magnetic-impurities~\cite{Rogachev2006} and flux pinning~\cite{Sato2022}. Quasiparticle-mediated cooling would give rise to hysteretic critical currents and should show enhancement of the critical current with in-plane magnetic fields. In contrast, the observed diffraction pattern in our work is symmetric in current, and we do not observe the enhancement of the critical current under applied in-plane magnetic field, as can be seen in Supporting Figure S7. Furthermore, the SAG nature of the wires is expected to allow for better thermalization of the Josephson junctions via substrate phonons than in the case of transferred wires~\cite{Murani2020}. Hence, quasiparticle cooling effects are unlikely to cause the non-convex diffraction pattern observed here. Magnetic impurities and flux vortex pinning effects, if present, should be equally effective for two- and three-terminal Josephson junctions made on the same chip using the same fabrication recipe. Yet, the characteristic non-convex MTJJ diffraction pattern from Figure~\ref{3tjj_map} is absent from all of the measured two-terminal devices. Hence, these effects are unlikely to be relevant here.

The addition of superconducting loops, split-gates, and more terminals can allow for custom tuning of the $0-\pi$ C$\varphi$R. Our work shows experimentally that MTJJs allow for the creation of Josephson devices with 0-$\pi$ C$\varphi$R without any magnetic elements in the junction area, along with gate tunability due to the semiconductor nature of the junction area. In addition, the observed few-quantum-mode transport at zero magnetic field coexisting with superconductivity demonstrates the potential of PbTe SAG nanowires for topological MTJJ devices and for studies of MZMs. Currently, the presence of accidental quantum dots may pose challenges for clean MZM experiments. We expect that further improvements in material growth could mitigate these challenges.

\section*{Data and Code Availability}
Source data for the figures presented in this paper and the data plotting code are available at the following Zenodo database \url{https://zenodo.org/records/13942725}.

\section*{Supporting Information}
Additional imaging data on nanowires and their synthesis details, additional data from two- and three-terminal Josephson junctions. Accidental quantum dot characterization is provided.
\bibliography{references}

\section*{Acknowledgements}
All aspects of the work at UMN were supported by the Department of Energy under Award No. DE-SC0019274. Portions of this work were conducted in the Minnesota Nano Center, which is supported by the National Science Foundation through the National Nano Coordinated Infrastructure Network (NNCI) under Award Number ECCS-1542202. Eindhoven University of Technology acknowledges the research program “Materials for the Quantum Age” (QuMat) for financial support. This program (registration number 024.005.006) is part of the Gravitation program financed by the Dutch Ministry of Education, Culture and Science (OCW). Eindhoven University of Technology acknowledges European Research Council (ERC TOCINA 834290).

\section*{Author Contributions}
M.G. and V.S.P. designed the experiments. M.G. and V.K. fabricated Devices 2,3 and 4 with help from L.S. M.G. performed the measurements on Devices 2, 3, and 4 and analyzed the data with help from V.K. G.M. and L.S. fabricated Device 1. C.R. performed measurements and analyzed data on Device 1 with help from G.M. P.L. and E.P.A.M.B. provided the nanowires. R.M. provided the theoretical model for the three-terminal device. All authors contributed to writing the manuscript. V.S.P. supervised the project.

\section*{Competing interests}
The authors declare no competing interests.
\end{document}


\section{Nanowire Growth}
\begin{figure}
    \centering
    \includegraphics[width=\textwidth]{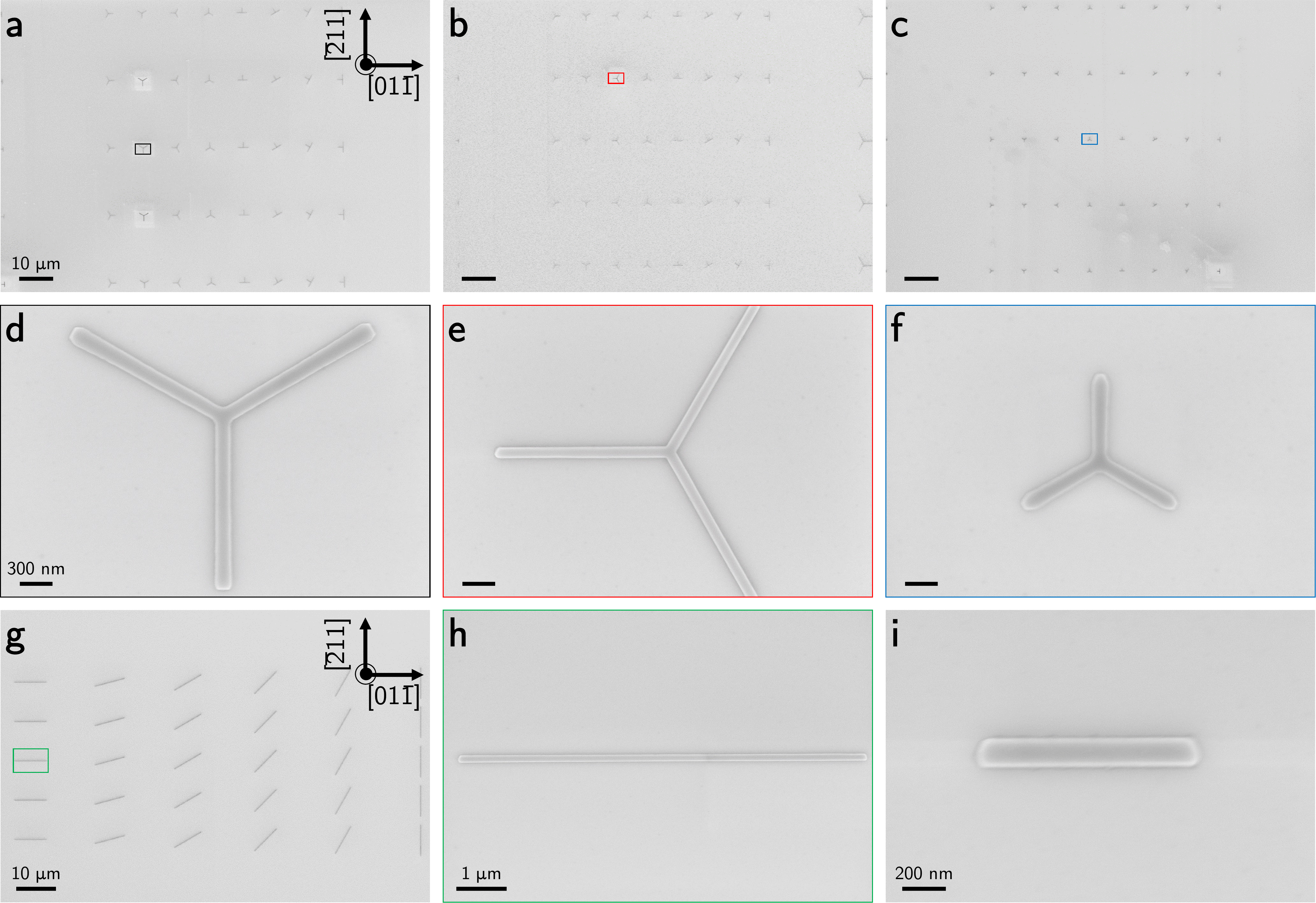}
    \caption{\textbf{SEM images of PbTe SAG on an InP(111)A}. Trijunctions intended for 3TJJs (\textbf{a-f}) with varying nanowire dimensions and orientations. Specifically, \textbf{a} and \textbf{d} feature trijunction with arms with a nominal length of 2 \textmu m and a width of 180 nm, while \textbf{b} and \textbf{e} showcase dimensions of 2 \textmu m length and 120 nm width. \textbf{c} and \textbf{f} present structures with dimensions of 500 nm length and 90 nm width. Additionally, \textbf{g-h} exhibit nanowires with a nominal length of 8 \textmu m and a width of 100 nm, while the nanowire in \textbf{i} features a length of 1 \textmu m and a width of 80 nm. Notably, no parasitic growth is observed, with growth confined exclusively to the predefined mask openings. Overall, a wide range of structures and dimensions for PbTe SAG can be obtained, allowing for tailoring to specific applications.}
    \label{sem_dev_si}
\end{figure}

PbTe nanowires were grown selectively on an InP(111)A substrate, following the method as outlined in Ref.~\cite{Fabrizio_Bakkers2022}. A SiN$\mathrm{_x}$ mask of 20nm was deposited on an InP(111)A wafer using plasma-enhanced chemical vapor deposition (PECVD). Nanowire patterns were lithographically defined and etched into the mask using a reactive-ion etching (RIE) using CHF$\mathrm{_3}$ with added O$\mathrm{_2}$. The residual resist left over from the lithography step is then chemically removed by ultrasonication in an acetone bath. The substrate is then shortly etched in a phosphoric acid solution ($\mathrm{H_2O}:\mathrm{H_3PO_4} = 10:1$) to remove native substrate oxides, after which it is rapidly transferred to the high-vacuum molecular beam epitaxy (MBE) system. Within the MBE system, the samples are degassed at 300\textdegree C for 1 hour. Subsequently, samples are annealed at 480\textdegree C under a Te overpressure. The temperature is then reduced to 340\textdegree C and growth commences under elemental fluxes of Te and Pb, which are 4E-7 mbar and 1.25E-7 mbar, respectively, as determined by a naked Bayard-Alpert ion gauge. The resulting height of the nanowires is $\sim$ 40nm as measured from the substrate. The resulting wire structures are displayed in Figure~\ref{sem_dev_si}.

\section{Additional Data for Device 1}
\begin{figure}
    \includegraphics[scale=1.8]{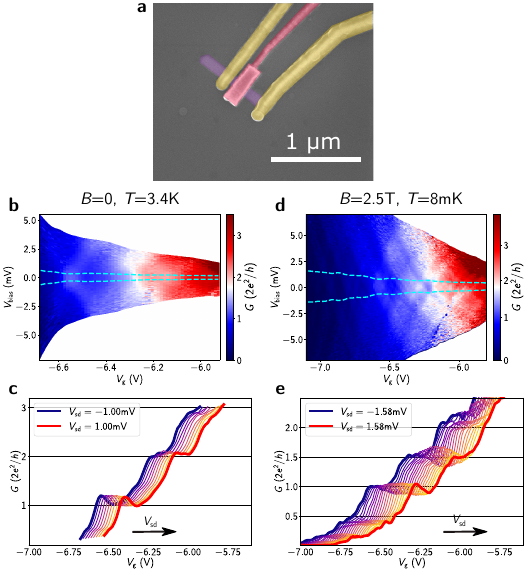}
    \caption{\textbf{Quantization with Ti/Au contacts.} \textbf{a} False-color scanning electron microscope (SEM) image of the device with Ti/Au contacts. Conductance as a function of source-drain bias $V_{\rm{bias}}$ and gate voltage $V_{\rm{g}}$ for Device 1 at \textbf{b} $B=0$T, $T=3.4$K and \textbf{d} $B=2.5$T, $T=8$mK. Conductance as a function of gate voltage for different $V_{\rm{sd}}$ for \textbf{c} $B=0$T, $T=3.4$K and \textbf{e} $B=2.5$T, $T=8$mK. The curves correspond to $V_{\rm{sd}}$ values between \textbf{c} $-1$ mV and $1$ mV and \textbf{e} $-1.58$ mV and $1.58$ mV (shown dashed cyan lines in panel \textbf{b,d}). The curves are each offset along the $V_{\rm{g}}$ axis for clarity with an arrow indicating the direction of increasing $V_{\rm{sd}}$.
}
    \label{Non_SC_QPC}
\end{figure}

The SEM image of the device is shown in Figure~\ref{Non_SC_QPC}\textbf{a}, the scattering region is defined by the top gate (shown in pink) with a length of $\sim 200$nm. We have performed measurements at high applied magnetic fields and elevated temperatures on Device 1. The obtained conductance maps show diamonds of uniform conductance and linecuts confirm these to be quantized conductance plateaus (Figure~\ref{Non_SC_QPC}), consistent with data presented in the main text. 

\section{Additional Data for Induced superconductivity}
\begin{figure}
    \includegraphics[width=\textwidth]{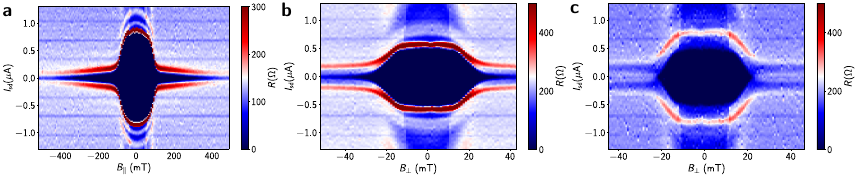}
    \caption{\textbf{Fraunhofer patterns.} \textbf{a} $R$ as a function of $I_{\rm{sd}}$ and in-plane magnetic field $B_{\parallel}$ for Device 2. $R$ as a function of $I_{\rm{sd}}$ and out-of-plane magnetic field $B_{\perp}$ for devices similar to Device 2 with wire diameter \textbf{b} 80nm \textbf{c} 60nm.}
    \label{wire_dia_IB}
\end{figure}

We present additional data on devices similar to the one presented in the main text, Device 2, but with different wire diameters. The observed diffraction patterns are consistent with one dimensional supercurrent flow as there is only one central superconducting lobe with suppression of the second lobe as the wire diameter decreases as seen in Figure~\ref{wire_dia_IB}. The magnitude of critical current also decreases as the wire diameter decreases. This shows good reproducibility of Josephson device fabrication on this material platform.

\section{Additional Data for Three-terminal Devices}
\begin{figure}
    \includegraphics[width=\textwidth]{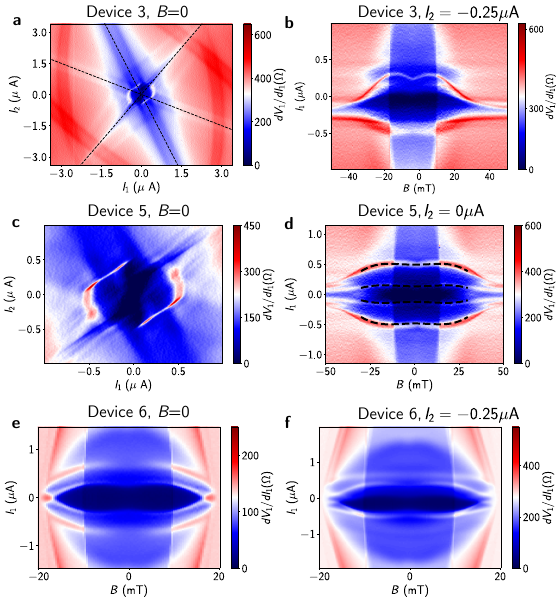}
    \caption{\textbf{Three-terminal transport.} $dV_1/dI_1$ as a function of bias currents $I_1$ and $I_2$ for \textbf{a} Device 3 \textbf{c} Device 5. Dashed black lines in \textbf{a} are drawn with approximate values in the $I_1-I_2$ plane. $dV_1/dI_1$ as a function of $I_1$ and $B$ \textbf{b} for $I_2=-0.25\mu$A for Device 3. \textbf{d} for $I_2=0\mu$A for Device 5. \textbf{e} for $I_2=0\mu$A for Device 6. \textbf{f} for $I_2=-0.25\mu$A for Device 6. The dashed black lines in \textbf{d} show the fit to the $\pi$-shifted quartet model.}
    \label{3tjj_map}
\end{figure}

We show the full differential resistance map for Device 3 in Figure~\ref{3tjj_map}\textbf{a}, showing the full superconducting arms in the $I_1$-$I_2$ plane, displaying the superconducting arms consistent with RCSJ network model. The lines with slope values are superimposed by dashed black lines. These lines give the resistance ratio values to be $R_1/R_2\sim3/2$, $R_2/R_3\sim2/3$ and $R_1/R_3\sim1$ in the RCSJ model discussed in Ref.~\cite{graziano2022}. The magnetic field diffraction pattern for $I_2=-0.25\mu$A is shown in Figure~\ref{3tjj_map}\textbf{b}. The observed pattern is opposite to what has been observed for $I_2=0.25\mu$A (Figure 3\textbf{d}). We also show data from a device similar to Device 3, labeled Device 5. Device 5 has a wire diameter of $100$nm compared to Device 3 which has a wire diameter of 60nm. Non-convex diffraction patterns and central superconducting features are also reproduced on this device. Another device from a different growth batch, Device 6, with a wire diameter of 100nm is also measured. No gates were fabricated on this device. The non-convex nature of the diffraction map is weaker in this device, possibly due to defects in the wire leading to suppression of $E_J'$, but is still clearly visible (Figure~\ref{3tjj_map}\textbf{e, f}).

We fit the non-convex feature present at $I_1\sim 0.5\mu$A in our experimental data and obtain the curve shown in Figure~\ref{3tjj_map}\textbf{d} by a dashed black line, with $E_J=0.29\frac{\hbar}{2e}\mu$A and $E'_J=-0.05\frac{\hbar}{2e}\mu$A and $A=0.023\mu\mathrm{m^2}$. This further validates our hypothesis that in planar geometry $E_J'$ are suppressed compared to the $E_J$ terms when the wire diameter is increased.

In-plane magnetic field measurement for Device 3 is shown in Figure~\ref{in_plane_nopi}, no field enhancement of critical current is observed, negating self-heating effects.

\section{Theoretical Model}
We derive the expression for critical current as a function of magnetic field using the model presented in Ref.~\cite{Melin2023}. The Josephson energy of a three-terminal device can be written as:
\begin{eqnarray} 
\begin{aligned}
    E=-E_J\cos{(\phi_1-\phi_0)}-E_J\cos{(\phi_2-\phi_0)}-E_J\cos{(\phi_1-\phi_2)}+\\
    - E'_J\cos{(\phi_1+\phi_0-2\phi_2)}-          E'_J\cos{(\phi_2+\phi_0-2\phi_1)}-E'_J\cos{(\phi_1+\phi_2-2\phi_0)}
    \label{cpr_mtjj}
    \end{aligned}
\end{eqnarray}

If the central area of the device is $A$ then for an applied field $B$ we can write:
\begin{eqnarray}
    \begin{aligned}
        \phi_0=-\frac{2\pi B A}{2\phi_e}=-\Phi/2\\
        \phi_1=\frac{2\pi B A}{2\phi_e}=\Phi/2\\
    \end{aligned}
\end{eqnarray}
Here we have used gauge invariance to set the phase offset between terminal 0 and 1 to be zero. We now arrive at Equation~\ref{simpl_JJe}:
\begin{eqnarray}
    \begin{aligned}
            E=-E_J\cos{(\Phi)}-2E_J\cos{(\Phi/2)}\cos{(\phi_2)}-E'_J\cos{(2\phi_2)}-2E'_J\cos{(3\Phi/2)}\cos{(\phi_2)}
    \end{aligned}
    \label{simpl_JJe}
\end{eqnarray}
To get an expression for the current-phase relation (C$\varphi$R) we perform derivative w.r.t the free phase variable $\phi_2$:

\begin{eqnarray}
    \begin{aligned}
    I(\Phi,\phi_2)=\frac{-2e}{\hbar}\frac{\partial E}{\partial\phi_2} = \frac{-2e}{\hbar}(2E_J\cos{(\Phi/2)}\sin{(\phi_2)}+\\4E'_J\cos{(\phi_2)}\sin{(\phi_2)}+2E'_J\cos{(3\Phi/2)}\sin{(\phi_2)})
    \end{aligned}
    \label{cphir}
\end{eqnarray}
Maximizing Equation~\ref{cphir} gives the expression for critical current for a given magnetic field. If $E'_J=0$, then Equation~\ref{cphir} has a maximum at $\phi_2=\pi/2$ and we recover the usual convex diffraction pattern. For small $E'_J$ we can expand the above expression near $\phi_2=\pi/2$, with $\phi_2=\pi/2+\epsilon$:

\begin{eqnarray}
    \begin{aligned}
      I(\phi,\epsilon)=\frac{-2e}{\hbar}(2E_J\cos{(\Phi/2)}-4E'_J\epsilon+2E'_J\cos{(3\Phi/2)})(1-\epsilon^2/2)
    \end{aligned}
    \label{Iphieps}
\end{eqnarray}
Here we have assumed $E'_j$ and $\epsilon$ to be small. Maximizing Equation~\ref{Iphieps} in $\epsilon$ we get the expression for the critical current as:
\begin{eqnarray}
    \begin{aligned}
        I_c(\Phi)=\frac{4e}{\hbar}(\lvert E_J \cos(\Phi/2) + E'_J \cos(3\Phi/2)\rvert)
    \end{aligned}
    \label{icexpr}
\end{eqnarray}
We have used Equation~\ref{icexpr} to fit the observed non-convex diffraction pattern.
\section{Quantum Dot Characterization with Ti/Al Contacts}
\begin{figure}
\includegraphics[width=\textwidth]{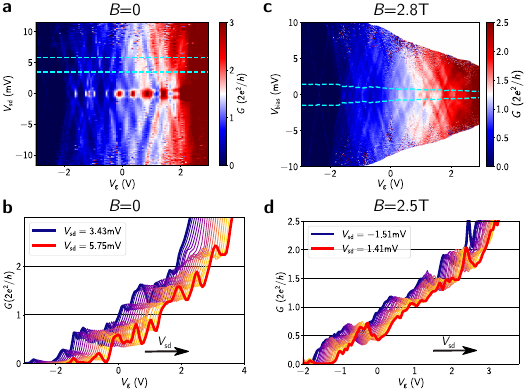}
    \caption{\textbf{Quantization in two-terminal Josephson device.} Conductance as a function of \textbf{a} $V_{\rm{sd}}$ and gate voltage $V_{\rm{g}}$ at $B=0$T \textbf{c} $V_{\rm{bias}}$ and gate voltage $V_{\rm{g}}$ at $B=2.8$T  . Conductance as a function of gate voltage for different $V_{\rm{sd}}$ for \textbf{b} $B=0$T and \textbf{d} $B=2.8$T. The curves correspond to $V_{\rm{sd}}$ values between \textbf{b} $3.43$ mV and $5.75$ mV and \textbf{d} $-1.51$ mV and $1.41$ mV (shown dashed cyan lines in panel \textbf{a,c}). The curves are each offset along the $V_{\rm{g}}$ axis for clarity with an arrow indicating the direction of increasing $V_{\rm{sd}}$.}
    \label{2t_jj_sc_qpc}
\end{figure}

In Figure~\ref{2t_jj_sc_qpc} we show few-mode transport in a two-terminal device. The observed plateau-like features and checkerboard-like resonances are observed in the conductance map in this device similar to Device 1 and 4. All three devices discussed in this work where pinch-off is observed for reasonable values of gate voltage have these resonances present in them likely due to an accidental quantum dot in the nanowire. Such dots are characterized in Device 4 in the range of gate voltage values corresponding to the near pinch-off of the device.

\begin{figure}
\includegraphics[width=\textwidth]{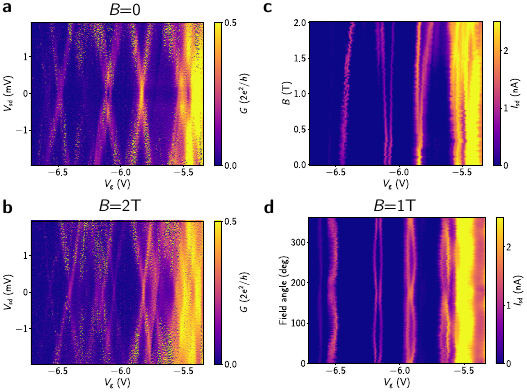}
    \caption{\textbf{Quantum dot characterization.} Conductance as a function of $V_{\rm{sd}}$ and gate voltage $V_{\rm{g}}$ at \textbf{a} $B=0$T \textbf{b} $B=2$T. $I_{\rm{sd}}$ as a function of \textbf{c} $B$ and $V_{\rm{g}}$ \textbf{d} magnetic field angle with the device plane and $V_{\rm{g}}$, magnitude of $B$ is 1T.}
    \label{3t_jj_dot}
\end{figure}

\begin{figure}
    \centering
    \includegraphics[width=0.8\textwidth]{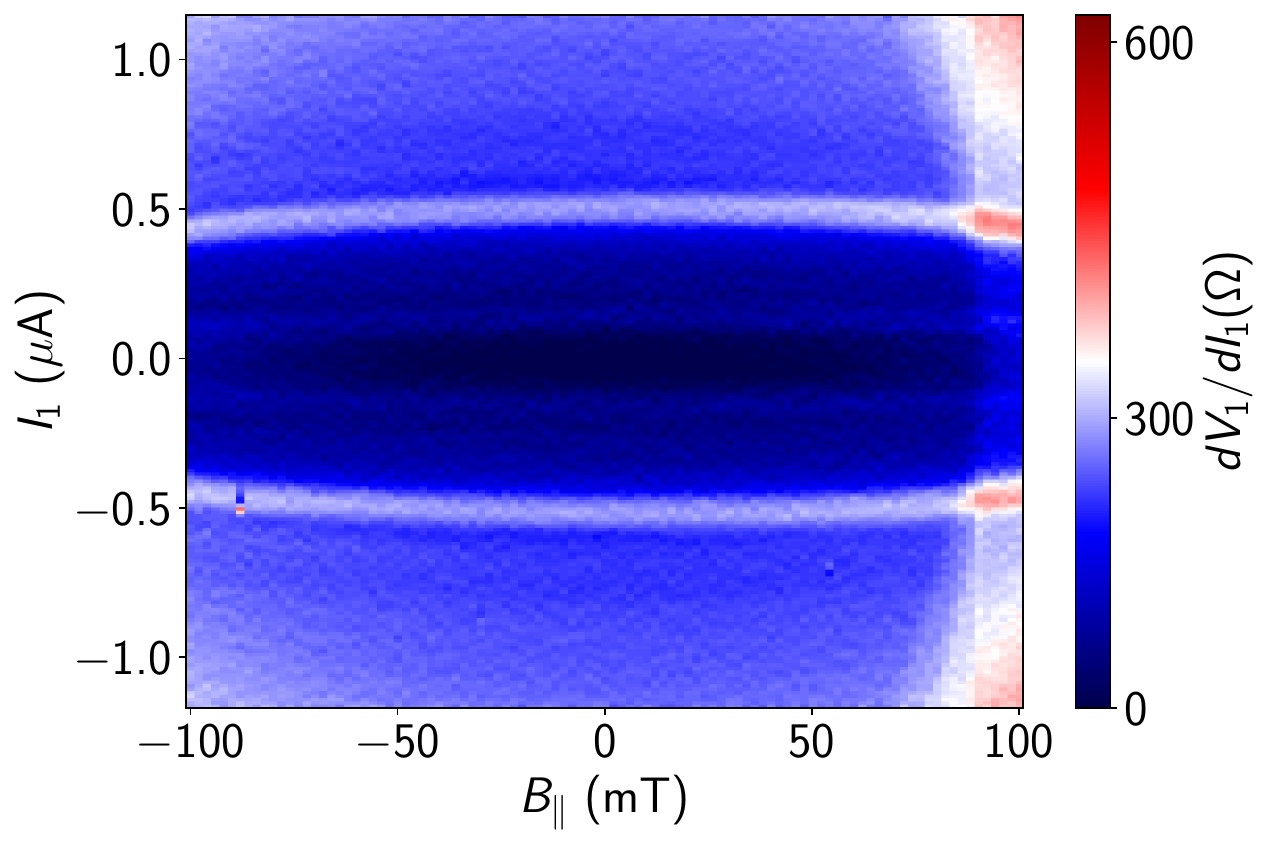}
    \caption{\textbf{In-plane magnetic field measurement.} $dV_1/dI_1$ as a function of $I_1$ and $B_\parallel$ for Device 3. }
    \label{in_plane_nopi}
\end{figure}

In Figure~\ref{3t_jj_dot}\textbf{a} we observe clearly defined Coulomb diamonds in a bias vs gate charge stability diagram at zero external magnetic field. Upon application of an out-of-plane 2T field, we observe the emergence of additional, smaller diamonds, as shown in Figure~\ref{3t_jj_dot}\textbf{b}. This emergence of additional diamonds with the addition of Zeeman energy, instead of widening of existing diamonds, is consistent with a vanishingly small charging energy of the accidental dot, as has been previously reported in PbTe nanowires~\cite{Gomanko2022}. In the absence of charging energy, the $g$-factor of the dot can be directly inferred from the height in bias voltage of the Zeeman split diamonds, as this height is precisely $E_{Z} = g\mu_BB$. As an example, the Zeeman-split diamond centered at $V_g \sim -5.8$V in  Figure~\ref{3t_jj_dot}\textbf{b} gives $E_z \sim 1.6$ meV, corresponding to $|g| \sim 14$, consistent with previous studies of PbTe quantum dots reported in Refs.~\cite{Gomanko2022,Fabrizio_Bakkers_gfactor_2022}.

Figure~\ref{3t_jj_dot}\textbf{c} shows the evolution of the quantum dot conductance peaks at a constant voltage bias of $V_{\rm{sd}}=100\mathrm{\mu}$eV as the out of plane magnetic field is swept from 0 to 2T. We observe that peaks that are degenerate or nearly degenerate at zero field split linearly upon application of field, as expected for spin degeneracy lifted by Zeeman energy in the absence of large charging energy. Additionally, by fixing the external field magnitude at $|B|=1$T and rotating the field in the plane normal to the wire, we see that this Zeeman splitting varies as a function of the angle of the applied magnetic field with minimal (maximal) splitting observed when the field is parallel (perpendicular) to the substrate, as shown in Figure~\ref{3t_jj_dot}\textbf{d}. This is indicative of a highly anisotropic $g$-factor in the dot, as has been reported in both fully gate-defined~\cite{Fabrizio_Bakkers_gfactor_2022} and geometrically-defined dots~\cite{Gomanko2022}. This is a consequence of a large Rashba-type spin-orbit coupling term tying the $g$-factor to the anisotropic confining fields present within the wire. As these devices were measured in a cryostat equipped with only a two-axis vector magnet, full determination of the $g$-factor tensor, as performed in Ref. \cite{Fabrizio_Bakkers_gfactor_2022} cannot be made for our accidental dots.

\section{Long Two-terminal Josephson Junctions}
\begin{figure}
    \centering
    \includegraphics[width=1\textwidth]{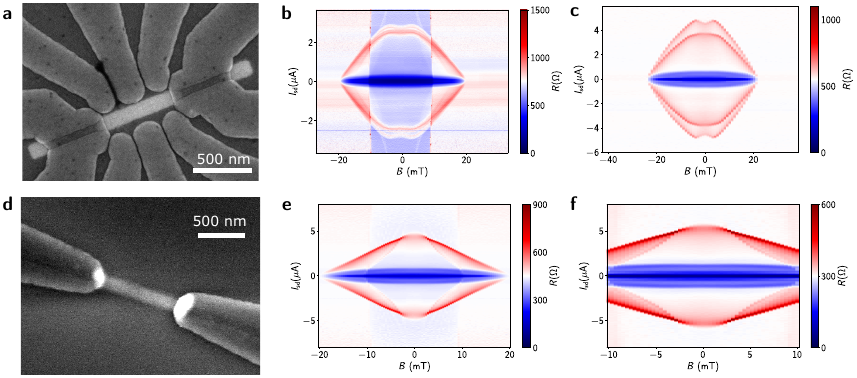}
    \caption{\textbf{Long two-terminal Josephson devices.} \textbf{a} SEM image of a device with Ti/Al contacts and side gates. $R$ as a function of $I_{\rm{sd}}$ and out-of-plane magnetic field $B$ for a similar device as shown in \textbf{a}; \textbf{b} with Ti/Al side gates and \textbf{c} with Ti/Au side gates. \textbf{d} SEM image of a device with Ti/Al contacts. $R$ as a function of $I_{\rm{sd}}$ and out-of-plane magnetic field $B$ for a similar device as shown in \textbf{d}; \textbf{e} with junction length of 740nm \textbf{f} with junction length of 600nm.}
    \label{long_2tjj}
\end{figure}

In a separate experiment we performed measurements on long two-terminal Josephson devices (contact spacing of $\sim 700$nm) from a different growth batch. In devices with multiple side gates (Figure~\ref{long_2tjj}\textbf{a}) the diffraction map shows a non-convex shape resonance outside the superconducting region (Figure~\ref{long_2tjj}\textbf{b, c}). The exact origin of these outer resonances in the resistive regime is unknown, but may be related to thermalization effects in the leads ~\cite{Ibabe2023} and may also potentially be due to the gate geometry as these resonances are not seen in the absence of local gates and with contact design similar to the one used in other two- and three-terminal devices (Figure~\ref{long_2tjj}\textbf{d, e, f}).

\section{Methods}

\subsection*{Device fabrication}
Electrodes were patterned using standard electron beam lithography (EBL) techniques using a tri-layer resist stack with two layers of 4 wt\% 495K poly(methyl methacrylate) (PMMA A4) and a single PMMA A2 layer. In-situ ion-mill etch was performed before the evaporation of superconducting Ti/Al ($5/50$ nm) or normal Ti/Au contacts in a UHV system. Approximately 40 nm of $\rm{Al_2O_3}$ dielectric was deposited using thermal atomic layer deposition (ALD). Using EBL, gates were defined over the junction area using the same resist stack that was used for contacts, and electrodes were deposited using electron-beam evaporation of Ti/Au (5 nm/80 nm).

\subsection*{Measurement details}
Data on Device 2-4 were obtained by low-noise DC  transport measurements in a $^3$He/$^4$He dilution refrigerator. For the conductance quantization data on Device 1, standard low-frequency lock-in techniques were used with a small excitation voltage and a frequency of 19 Hz. For all the voltage bias measurements, the raw data is corrected by subtracting the series filter and the ammeter resistances ($\sim7.8\rm{k}\Omega$ for Device 1, $\sim9.7\rm{k}\Omega$ for Device 4 and gate sweeps for Device 2, and $\sim8.8\rm{k}\Omega$ for the magnetic field sweep for Device 2). Low-pass Gaussian filtering was used to smooth numerical derivatives. We have subtracted a contact resistance of $6.2\rm{k}\Omega$ for Device 1, $100\Omega$ for Device 2, and $120\Omega$ for the leg biased with $I_1$ and $132\Omega$ for the leg biased with $I_2$ for Device 3. For Figure 3\textbf{b}, we have also subtracted an instrument offset of $44\mu$V. No subtraction of contact resistance is done for Device 4. 

\bibliography{references}
\newpage

